\documentclass[conference]{IEEEtran}
\usepackage{amsmath}
\usepackage{graphicx}
\usepackage{color,soul}

\newcommand{\tabincell}[2]{\begin{tabular}{@{}#1@{}}#2\end{tabular}}

\ifCLASSINFOpdf
\else
\fi
\begin{document}
%
\title{A Closed-Loop UL Power Control Scheme for Interference Mitigation in Dynamic TD-LTE Systems}
%
\author{\IEEEauthorblockN{Qinqin~Chen,~Hui~Zhao,~Lin~Li,~Hang~Long,~Jianquan~Wang,~Xiaoyue~Hou}
\IEEEauthorblockA{Key Laboratory of Universal Wireless Communication, Ministry of Education\\
Wireless Signal Processing \& Network Lab, BUPT\\
Beijing, China\\
Email: qinchen@bupt.edu.cn}
}
\maketitle
\begin{abstract}
The TD-LTE system is envisaged to adopt dynamic time division duplexing (TDD) transmissions for small cells to adapt their communication service to the fast variation of downlink (DL) and uplink (UL) traffic demands. However, different DL/UL directions for the same subframe in adjacent cells will result in new destructive interference components, i.e., eNB-to-eNB and UE-to-UE, with levels that can significantly differ from one subframe to another. In this paper, a feasible UL power control mechanism is proposed to manage eNB-to-eNB interference, where different UL power control parameters are set based on different interference level. We consider the geometric location information and the subframe set selection process about adjacent eNBs when the interference level is estimated. The performance of the proposed scheme is evaluated through system level simulations and it is shown that the scheme can achieve preferable improvement in terms of UL average and 5\%-ile packet throughputs compared with the original scheme without power control. Also, the UE-to-UE interference is not worse when the UE transmit power become higher.
\end{abstract}
\begin{IEEEkeywords}
Dynamic TDD, Interference mitigation, UL power control, TD-LTE.
\end{IEEEkeywords}
\IEEEpeerreviewmaketitle
\section{Introduction}
\IEEEPARstart{T}{he}
future wireless communication systems should support various multimedia services, such as voice over IP (VoIP), video streaming, interactive gaming and peer-to-peer (P2P) file transfer, etc. Because of these various multimedia services, the traffic asymmetry property will be very remarkable. It can be envisaged that in the networks, e.g., the 3GPP LTE Release 12/13 networks, small cells will prioritize time division duplexing (TDD) schemes over frequency division duplexing (FDD) ones since TDD transmissions are particularly suitable for hot spot scenarios with traffic fluctuations in both link directions [1]. In normal TDD system, the DL/UL slot assignment is fixed and aligned among the neighboring cells, which is the so-called static and synchronized TDD. In this line, a new technology has recently emerged, referred to as dynamic TDD, in which TDD DL and UL subframes can be dynamically configured in small cells to adapt their communication service to the fast variation of DL/UL traffic demands in either direction. The application of dynamic TDD in homogeneous small cell networks has been investigated in recent works with positive results [2].
\par In the dynamic TDD networks, the configuration of the DL/UL resource can be done separately for each cell. Thereafter the main practical technical challenge is the emergence of new types of inter-cell interference such as interference between neighboring base stations (eNB-to-eNB) and between user terminals (UE-to-UE). The eNB-eNB interference is especially detrimental and has been shown to significantly impact UL performance[3].
To mitigate the interference of the LTE or LTE-A networks, some schemes have been studied[4][5].
\par To alleviate the interference problems (eNB-to-eNB or UE-to-UE interference) of the dynamic TDD system, several schemes were proposed. In [6], the authors study cell clustering based techniques. The ability of a base station to measure the signal from another base station is hence a key ingredient of such a method.  Soft Frequency Reuse (SFR) schemes were investigated under LTE framework to provide a higher rate to disadvantaged cell-edge UEs [7][8]. SFR, which utilizes Resource Blocks (RB) allocation in cluster cells, is an effective method to mitigate inter-cell interference in downlink service system. However, for the system with both uplink and downlink loads, its spectrum efficiency is low because the outward users close to the cell edge in a cell cannot utilize the all frequency resource. In [9], the authors investigate an inter-cell coordination scheme that coordinates the transmission time and mode of users in neighboring cells.
\par Since neighboring cells can select DL/UL configuration separately and ecperience different interference in different subframes, some power control schemes arise. Each eNB adjusts the transmit power in some specific DL subframes which may cause severe eNB-to-eNB interference to the neighboring cells[10]. Also, two power control schemes are studied, one in which the UL subframe sets are configured statically and the other in which the UL subframe sets are configured dynamically and are changed according to the continuously monitored interference level [11]. In the above power control schemes, the reconfiguration information of neighboring cells in the actual scenario can not be considered. In this paper, we focus on UE transmit power adjustments process based on the interference level and its performance in combatting the strong eNB-to-eNB interference. In particular, we consider the geometric location information and the subframe set selection process of interfering eNBs to measure the interference level.
\par The rest of the paper is organized as follows: network scenario information is provided in Section II. The proposed UL power control scheme is presented in Section III. Simulation results and some concluding remarks are provided in Section V and VI , respectively.

\section{NETWORK SCENARIO}
\subsection{ Deployment Scenarios and Interference Types}
The work in LTE Release 12 on dynamic TDD reconfiguration builds on Release 11 studies, both known as enhanced interference mitigation and traffic adaptation (eIMTA). Various heterogeneous scenarios (pico cells only, macro-pico cells, macro-femto cells) with co-channel and adjacent channel deployments were considered for eIMTA [1]. Of these scenarios, multiple outdoor pico cells with and without macro coverage were selected as the best candidates and the following two scenarios were given highest priority: \par$\bullet$ multiple outdoor co-channel pico cells
\par$\bullet$ and multiple outdoor co-channel pico cells with adjacent channel macro cells.
\par It is observed that for the case of very low adjacent channel interference ratio, the second scenario essentially reduces to two independent deployments, one consisting of just macro cells and the other of just pico cells. In this paper, we adopt this scenario just focus on the types of interference in pico cells.

\begin{figure}[t]
\centering
\includegraphics[width=0.45\textwidth]{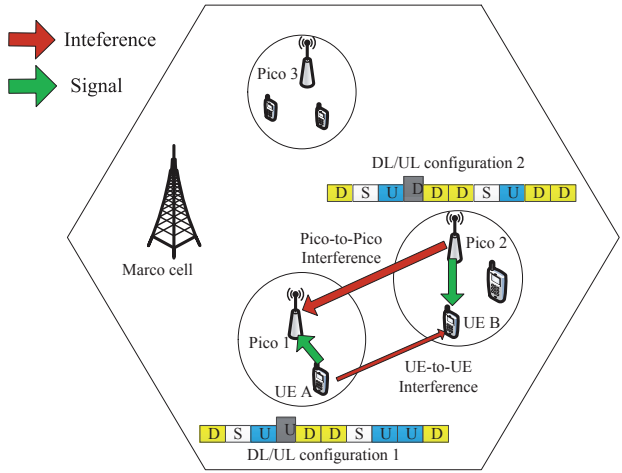}
\vspace*{-5pt}
\caption{The types of interference in pico cells of dynamic TD-LTE system.}
\label{1}
\vspace*{-15pt}
\end{figure}

\par Fig.1 shows a heterogeneous network deployment with multiple TDD pico cells in the macro cell area and highlights the eNB-to-eNB and UE-to-UE interference encountered in a pico cell based TD-LTE system assuming the macro cell interference can be ignored. Pico1 cell and Pico2 cell have different configurations. When the cells transmit in the fourth subframe, eNB2 will interfere eNB1 in transmitting mode. Similarly, UE A will also interfere UE B in receiving mode.

\subsection{Uplink Power Control in LTE}
UL power control in LTE, which is one of the mechanisms used for combatting interference, comprises both open and closed loop components as well as adjustments based on transmission parameters. The open loop component uses the target received power at the cell, $P_0$, and an estimate of the DL path loss weighted by a compensation factor, $\alpha$, to adjust the terminal power where $P_0$ and $\alpha$ are provided by the eNB. The closed loop component adjusts the terminal transmit power via explicitly transmitted power control commands. In a subframe, power is set on a channel basis per component carrier (in the case of carrier aggregation) and then reduced if the UE maximum power would be exceeded by the sum of the channel power. Details of UL power control can be found in Section 5.1 in [12].
\par In LTE Release 12, subframe set based power control has been introduced whereby each of two sets of subframes can have its own set of open loop power control parameters ($P_0$, $\alpha$) and separate transmit power control commands (TPC). Determination of power in a subframe is based on the open loop and closed loop components that apply to the subframe set to which the subframe belongs.

\section{PROPOSED CLOSED-LOOP UL POWER CONTROL OPERATION}
\subsection{Basic Idea}
\begin{figure}[b]
\centering
\includegraphics[width=0.45\textwidth]{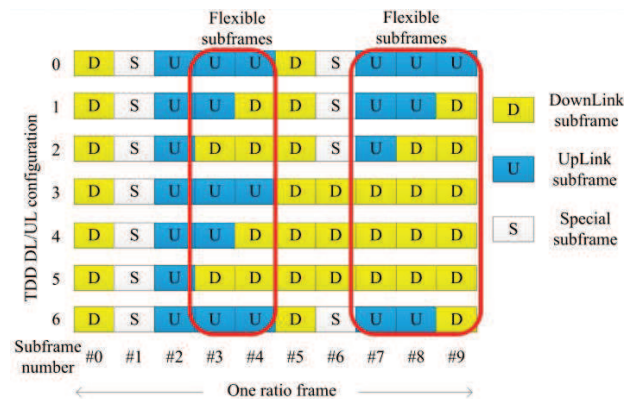}
\vspace*{-5pt}
\caption{Dynamic TDD DL/UL configurations of TD-LTE system.}
\label{2}
\vspace*{-10pt}
\end{figure}
Dynamic TD-LTE system has 7 different DL/UL configurations as shown in Fig.2. And eNBs select corresponding configurations to adapt the traffic. Configuration 0 will not cause eNB-to-eNB interference to any other configurations. For configuration 1-6, since the special subframe is usually used as downlink subframe, therefore, there are no eNB-to-eNB and UE-to-UE interference in subframe $\sharp$0, $\sharp$1, $\sharp$2, $\sharp$5 and $\sharp$6. This paper only needs to consider the possible interference in subframe $\sharp$3, $\sharp$4, $\sharp$7, $\sharp$8 and $\sharp$9. For conveniences, subframe $\sharp$0, $\sharp$1, $\sharp$2, $\sharp$5 and $\sharp$6 are referred as FIS (Fixed subframe) and the other subframes are called as FLS (Flexible subframe).

\par In order to improve the quality of uplink, a feasible method is to increase the received signal SINR in the subframes which suffer from the eNB-to-eNB interference. The SINR of eNB $i$ is

\begin{equation}\label{1}
    SINR_i=\frac{P^{UEsignal}_{i}}{P^{CCI}_{i}+P^{N_{0}}_{i}},
\end{equation}

where $P^{UEsignal}_{i}$ is the effective UE transmission power which eNB $i$ receives, $P^{N_{0}}_{i}$ is white Gaussian noise and $P^{CCI}_{i}$ is the received interference signal power from adjacent eNBs. Then we have

\begin{equation}\label{2}
    P^{CCI}_{i}=\sum_{k=1}^{N} P_{i,k}^{CCI}=\sum_{k=1}^{N} (P_{i,k}^{interf}PL_{i,k}),
\end{equation}

Assume $N$ is the number of interfering eNBs, $P_{i,k}^{CCI}$ is the received interference signal power from eNB $k$, $P_{i,k}^{interf}$ is the transmission power of eNB $k$ and $PL_{i,k}$ is the path loss between eNB $i$ and the interfering eNB $k$.

\par For increasing the SINR, a method is to enhance $P^{UEsignal}_{i}$ and another is to reduce $P_{i,k}^{interf}$. However, LTE systems should avoid changing eNBs' transmission power frequently. Therefore, in the proposed scheme, we achieve this goal by increasing $P^{UEsignal}_{i}$ properly. The appropriate UE transmit power $P^{UEsignal}_{i}$ and $P^{UEsignal}_{i}+\Delta_{set}$ ($\Delta_{set}= \Delta_{set,1}, \Delta_{set,2}, \ldots, \Delta_{set,max}$)are set to small cells for FIS and FLS, respectively. The interference level is different in different subframes of neighboring cells, thus $\Delta_{set}$ should changes synchronously. If the neighboring cells adopt different DL/UL configurations, the interference depends on the flexible UL subframe itself and it may be, for example, different in subframe $\sharp$3 than in $\sharp$8. Although subframe $\sharp$3 and $\sharp$8 belong to FLS.

\par In this proposed scheme, the first step of the power adjustment is to estimate the number of the interfering eNBs, and each eNB sends the UL configuration information to the interfered ones. Then the interfered eNBs will calculate the interference level and the UEs carry out power control process based on the indicator to mitigate its interference. These steps will be also done periodically. The details of each step will be given in the following sections.

\subsection{Interfering eNB Set}
Due to the different distance and propagation environment, the interference levels of eNBs are also different and we only regard the strong interfering ones as interference sources. The path loss between eNBs can be used to characterize the strength of the interference. Then each eNB measures the path loss of the reference signal received power from its neighboring eNBs and chooses $N$ interfering eNBs reasonably.
\par The received interference signal power from eNB $k$ will be

\begin{equation}\label{3}
    P_{i,k}^{CCI}=P_{i,k}^{interf}PL_{i,k}\propto\frac{P_{i,k}^{interf}}{d_{i,k}^{\alpha}},
\end{equation}

where $d_{i,k}^{\alpha}$ is the distance between eNB $i$ and eNB $k$, $\alpha$ is path loss coefficient, and $PL_{i,k}$ is the path loss. When the propagation environment is identical, $\alpha$ is also equal. In homogeneous networks, all eNBs' transmit power are equivalent, and we can simply determine whether the eNBs should be regard as interference sources by the geometrical location. The ones with shorter distances will be more likely classified as interfering eNB. In heterogeneous network, $P_{i,k}^{interf}$ is different and the interfering eNB set will be determined based on $d_{i,k}^{\alpha}$ and $P_{i,k}^{interf}$. For different types of eNBs, the transmit power are known, e.g. macro eNBs usually transmit with 46 dBm power, and small cell eNBs transmit with 30 dBm or 24 dBm power, etc.
\par In the scenario of just Pico cells, the eNBs are randomly distributed. Therefore we distinguish the boundary of the interference sources by setting the threshold $P_{threshold}$. When $PL_{i,k} \leq P_{threshold}$, the ones belong to the strong interference resources and when $PL_{i,k}>P_{threshold}$, we will not consider the impact of these eNBs.

\subsection{Configurations Information Exchange Between eNBs}
After the number of interference sources is determined, their configuration information are sent to the interfered eNB. This information can be sent in two different ways. A straightforward way is to send a bitmap with 5 bits indicating the link information of FLS. If it is DL subframe, 1 is sent, otherwise, 0 will be sent. Take the DL/UL configuration 1 as an example, the bitmap will be 01001. Another way is to send the DL/UL configuration number. Therefore, totally only 3 bits are needed. It will send 001 when eNB chooses configuration 1. When 001 is received, the configuration information (01001) will be generated. Considering the command burden, this paper chooses the latter one.
Each eNB chooses corresponding configuration to adapt the small cell traffic periodically. Thus the power control period follows the DL/UL reconfiguration period, e.g. 10ms, 200, 640ms, etc.

\subsection{Interference Level Model}
To estimate the interference level more precisely, a interference level indicator $I$ which takes into account all major factors can be defined. The received signal at interfered eNB $i$ is written as
\begin{equation}\label{4}
    \textbf{y}_{i}=\textbf{H}_{i,j}^{U}\textbf{x}_{j}+\sum^{N}_{k=1} \alpha_{k}\textbf{H}_{i,k}^{B}\textbf{s}_{k}+\textbf{IF}_{i,other}+\textbf{n}_{i},
\end{equation}
Where $\textbf{y}_{i}$ is the received signal at eNB $i$, $\textbf{H}_{i,j}^{U}$denotes the channel only considers the large scale fading between UE $j$ and eNB $i$, $\textbf{x}_{j}$ is the UL transmit signal from UE $j$ in eNB $i$. $N$ is the number of interfering eNBs.$\alpha_{k}$ is one of the five bits which denotes the configuration information of the interfering eNB $k$. $\textbf{H}_{i,k}^{B}$ is the channel only consider large scale fading between eNB $i$ and eNB $k$ and $\textbf{s}_{k}$ is the transmit signal of eNB $k$. $\textbf{IF}_{i,other}$ denotes other interference signals. $\textbf{n}_{i}$ is the complex Gaussian noise signal, and its power is $N_0$. Therefore, we have
\begin{equation}\label{5}
    SINR_{i}=\frac{\parallel\textbf{H}_{i,j}^{U}\textbf{x}_{j}\parallel_{F}^{2}}
    {\sum_{k=1}^{N}\parallel\alpha_{k}\textbf{H}_{i,k}\textbf{s}_{k}\parallel_{F}^{2}
    +\parallel\textbf{IF}_{i,other}\parallel_{F}^{2}+N_{0}},
\end{equation}
Assume the UE $j$ transmit power is $P_{i,j}^{U}$, the eNB $k$ transmit power is $P_{i,k}^{B}$
and the path loss between eNB $i$ and eNB $k$ is $PL_{i,k}^{B}$. Also, $\textbf{IF}_{i,other}$
is not considered. The above $SINR_{i}$ can be rewritten as
\begin{equation}\label{6}
    SINR_{i}=\frac{P_{i,j}^{U}PL_{i,j}^{U}G_{i}g_{i,j}}{\sum_{k=1}^{N}\alpha_{k}
    P_{i,k}^{B}PL^{B}_{i,k}G_{i}G_{k}+N_{0}},
\end{equation}
where $G_i$ is the antenna gain of eNB $i$, $G_k$ is the antenna gain of eNB $k$ and $g_{i,j}$ is the antenna gain of UE $j$. The UL capacity in eNB $i$ is

\begin{equation}\label{7}
    C_{i}=B\log_{2}(1+SINR_{i}),
\end{equation}
Since the eNB-to-eNB interference is the major interference, so it is reasonable to assume that $SINR_{i}\gg1$ after adjust the UE transmit power. Then the capacity can be written as

\begin {equation}\label{8}
\begin{array}{l}
\begin {aligned}
 C_{i} &\approx B[\log_{2}(P_{i,j}^{U}PL_{i,j}^{U}G_{i}g_{i,j})\\
 &- \log_{2}(\sum_{k=1}^{N}\alpha_{k}P_{i,k}^{B}PL_{i,k}^{B}G_{i}G_{k}+N_{0})],
\end {aligned}
\end{array}
\end {equation}
where $G_i$ , $G_k$ and $N_0$ are constant and their effects can be neglected. According to the above formula, the path loss $PL_{i,k}^{B}$ and the adjacent eNB transmit power $P_{i,k}^{B}$ are primary influencing factors. Therefore, we propose to define the interference level indicator $I$ in FIS subframes for eNB $i$ as below

\begin{equation}\label{9}
    I=\log_{2}(\sum_{k=1}^{N}\alpha_{k}P_{i,k}^{B}PL_{i,k}^{B}),
\end{equation}
\par When the $SINR_{i}$ is very low, the interfering signal is a key factor in the eNB received signal. So, $\sum_{k=1}^{N}\alpha_{k}P_{i,k}^{B}PL_{i,k}^{B}G_{i}G_{k}\gg P_{i,j}^{U}PL_{i,j}^{U}G_{i}g_{i,j}$, it is also reasonable that the interference level indictor is defined as formula(9).
\par In the process of the proposed scheme, the interfering eNBs' configuration information is received and the interference level indicator $I$ will be calculated. According to the $I$, UEs adjust the transmission power where $\Delta_{set}$ follows the rule of Table I. When $\alpha_k=1(k=1, 2, \ldots, N)$, the max interference level indicator $I_{max}$ of eNB is $\log_{2}(\sum_{k=1}^{N} P_{i,k}^{B}PL_{i,k}^{B})$.

\begin{table}[b]
  \centering
  \caption{THE RELATIONSHIP OF $I$ AND $\Delta_{set}$}\label{4}
   \begin{tabular}{|c|c|c|c|c|c|c|}
   \hline
   $I$ & [0,$I_{1}$] & [$I_{1}$,$I_{2}$] &\ldots& [$I_{i-1}$,$I_{i}$] &\ldots& [$I_{n}$,$I_{max}$] \\
   \hline
   $\Delta_{set}$ & $\Delta_{set,1}$ & $\Delta_{set,2}$ &\ldots& $\Delta_{set,i}$ & \ldots& $\Delta_{set,max}$ \\
   \hline
 \end{tabular}
\end{table}

\begin{table}[t]
\centering
  \caption{THE VALUE OF $I$ AND $\Delta_{set}$}\label{5}
   \begin{tabular}{|c|c|c|}
   \hline
   Parameter & $I_i$  &$\Delta_{set}$(dBm) \\
   \hline
   ($I_{1}$,$\Delta_{set,1}$) & $1/3I_{max}$ & $0$\\
   \hline
   ($I_{2}$,$\Delta_{set,2}$) & $1/2I_{max}$ & $1$\\
   \hline
  ($I_{3}$,$\Delta_{set,3}$) & $2/3I_{max}$ & $3$\\
   \hline
   ($I_{max}$,$\Delta_{set,4}$) & $I_{max}$ & $5$\\
   \hline
 \end{tabular}

\end{table}
\begin{table}[t]
  \centering
  \caption{BASIC SIMULATION PARAMETERS}\label{6}
  \begin{tabular}{|l|l|}
    \hline
    Parameter & Assumptions used for evaluation \\
    \hline
    Carrier frequency & 2GHz \\
    \hline
    Bandwidth & 10MHz \\
    \hline
    ISD & 500m \\
    \hline
    Outdoor Pico deployment & 40m radius, random deployment \\
    \hline
    Max power & 24dBm (Pico)/23 dBm (UE) \\
    \hline
    Pico Antenna & 5dBi, 1TX, 2RX \\
    \hline
    UE Antenna & 0dBi, 1TX, 2RX \\
    \hline
    Noise power & -174dBm/Hz \\
    \hline
    Outdoor Pico to outdoor Pico & Small scale fading is not modeled \\
    \hline
    Scheduling & First-in first-out with proportional fair \\
    \hline
    FTP Traffic model & \tabincell{l} {Packet size 0.5MB\\DL arrival rate\\ -$\lambda_{D}$ = ${0.25, 0.5, 0.75, \ldots, 2.0}$\\UL arrival rate\\-$\lambda_{U}$ = $\lambda_{D}/2$}\\

    \hline
     \tabincell{l} {UL power control parameter\\ and assumption} & \tabincell{l} { $\alpha$ = 0.8\\$P_{0}$ = -76 dBm \\$P_{threshold}$ = 130dB}\\

    \hline
  \end{tabular}
\end{table}
\section{SIMULATIONS AND DISCUSSION}
\subsection{Simulation Parameters}
This paper adopts DL/UL packet throughput as the performance metric. The average packet throughput provides a measure of overall system performance, whereas the 5\%-ile packet throughput is used to assess the cell-edge performance. We assume that the hypothetical value of $I_i$ and $\Delta_{set,i}$ are listed in Table II. $I_i$ is set three demarcation points and four regions. When the UE transmit power is adjusted, it is not greater than the maximum value. If the subframe is FIS, the UE transmit power don't make change. The dynamic TDD DL/UL reconfiguration period and the power control period are selected as 10 msec.

\par In the simulation, there are 19 three-sector macro sites and 4 picos are randomly deployed in each sector. 10 UEs are uniformly distributed inside each pico cell. Also, the macros are not active and each pico can change TDD DL/UL configuration dynamically. Table III provides other system level simulation parameters specific to our study. We use a non-full buffer traffic model to generate traffic in the system. Specifically, we use the FTP traffic mode characterized by poisson distributed arrivals with an arrival rate of $\lambda_{D}$(for DL arrivals), and $\lambda_{U}$(for UL arrivals) and adopt a fixed file size of 0.5MB per packet.

\begin{figure}[t]
\centering
\includegraphics[width=0.45\textwidth]{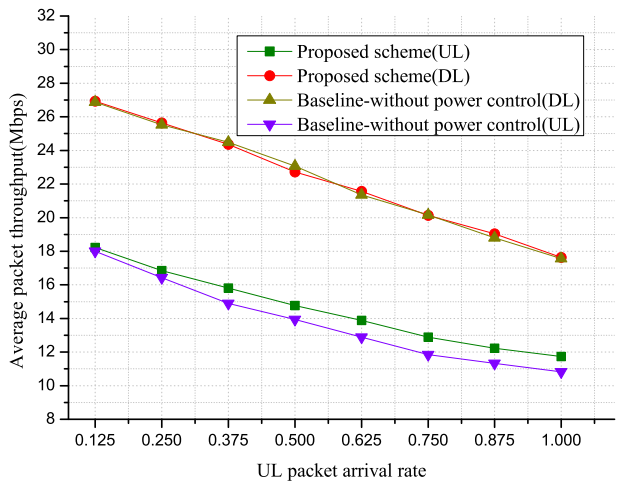}
\vspace*{-5pt}
\caption{Average DL/UL packet throughput vs. UL packet arrival rate.}
\label{3}
\vspace*{-5pt}
\end{figure}

\begin{figure}[t]
\centering
\includegraphics[width=0.45\textwidth]{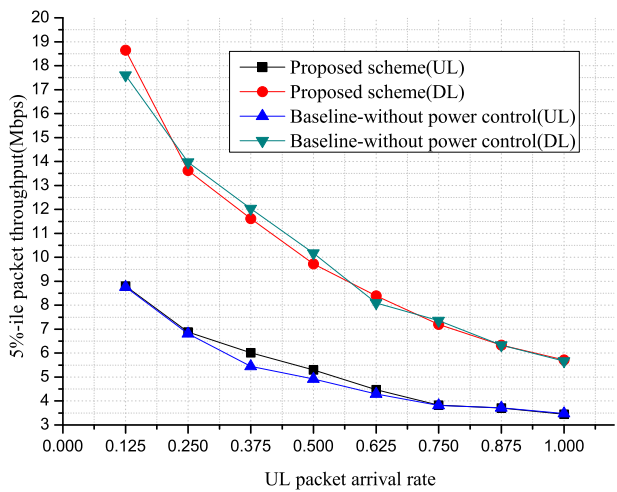}
\vspace*{-5pt}
\caption{5\%-ile DL/UL packet throughput vs. UL packet arrival rate.}
\label{4}
\vspace*{-5pt}
\end{figure}

\subsection{Results}
Fig.3 and Fig.4 show the average DL/UL packet throughput and the 5\%-ile packet throughput for the proposed scheme vs. the baseline, respectively. At very low packet arrival rates ($\lambda_U\leq0.25$), the experienced interference as well as the probabilities of reconfiguration are extremely low. This results in low gain of the proposed scheme. The average UL packet throughput increases less than 2\%, meanwhile the cell-edge throughput obtain little improvement.
As the UL packet arrival rate becomes higher, there is an increase in the number of active downloads and/or uploads. This has the twofold effect: \par$\bullet$ increasing the experienced interference;\par$\bullet$ and allowing for frequent changes in interference for the UL subframes.
\par The proposed scheme monitors the surrounding interference and sets appropriate $\triangle_{set}$. As a result, we observe that the average UL packet throughput gains over 5\% ($\lambda_U>0.25$). And from Fig.4, the remarkable performance gains are observed for low and medium traffic loadings ($0.25<\lambda_U<0.75$). As $\lambda_U\longrightarrow1.0$, the interference increases and the cell-edge UEs are interfered more seriously. The 5\%-ile packet throughput gains nothing, but the average UL packet throughput still get the gain over 8\%.
\par On the one hand, the UEs of higher transmission power interfere the adjacent edge UEs more seriously(UE-to-UE interference) and result in lower DL packet throughput. On the other hand, the UL channel quality will be better and the data packets are transmitted more efficiency. Accordingly, the UL time slot resource is reduced relatively and it is opposite for the DL time slot resource. This factor will be beneficial to DL packet throughput. Due to the interaction of the two factors, Fig.3 shows that the average DL packet throughput has almost no significant changes. The 5\%-ile DL packet throughput becomes lower slightly ($0.25<\lambda_U<0.5$) from Fig.4.

\section{CONCLUSION}
 In this paper, we have investigated a closed-loop UL power control scheme to combat the severe eNB-to-eNB interference that results from independent dynamic TDD DL/UL reconfiguration. In this scenario which just focus on the types of interference in pico cells, we propose an algorithm to estimate the eNB-to-eNB interference level. From the simulation results, we can conclude that the significant performance gains can be achieved for the average UL packet throughput. And the 5\%-ile packet throughput gains over 9\% for medium data traffic. When the traffic load is higher, the parameters($I_{i}$ and $\triangle_{set}$) need to be considered more accurately. Meanwhile, the UE-to-UE interference is not as serious as we expected when the UEs increase the transmit power.

\section*{Acknowledgment}
This work is supported in part by the National Key Technology R\&D Program of China under Grant 2014ZX03003011-004, the China Natural Science Funding (No.61302088) and the Research Fund for the Doctoral Program of Higher Education under Grant No. 20130005120003.


%

\end{document}